\documentclass[letterpaper, 10 pt, conference]{ieeeconf}  

\IEEEoverridecommandlockouts                              

\usepackage{graphicx}
\usepackage[caption=false, font=footnotesize]{subfig}

\title{\LARGE \bf
Analysing Explanation-Related Interactions in Collaborative Perception-Cognition-Communication-Action
}

\author{Marc Roig Vilamala$^{1}$, Jack Furby$^{1}$, Julian de Gortari Briseno$^{2}$, Mani Srivastava$^{2}$, Alun Preece$^{1}$ and \\ Carolina Fuentes Toro$^{1}$  
\thanks{$^{1}$Cardiff University, UK. Emails:
        {\tt\small \{RoigVilamalaM, furbyjl, preecead, fuentestoroc\}@cardiff.ac.uk}}%
\thanks{$^{2}$University of California, Los Angeles, USA
        {\tt\small \{julian700, mbs\}@ucla.edu}}%
}

\begin{document}

\maketitle
\thispagestyle{empty}
\pagestyle{empty}

\begin{abstract}

Effective communication is essential in collaborative tasks, so AI-equipped robots working alongside humans need to be able to explain their behaviour in order to cooperate effectively and earn trust. We analyse and classify communications among human participants collaborating to complete a simulated emergency response task. The analysis identifies messages that relate to various kinds of interactive explanations identified in the explainable AI literature. This allows us to understand what type of explanations humans expect from their teammates in such settings, and thus where AI-equipped robots most need explanation capabilities. We find that most explanation-related messages seek clarification in the decisions or actions taken. We also confirm that messages have an impact on the performance of our simulated task.

\end{abstract}


\section{INTRODUCTION}

Effective human-robot teaming is seen as a key enabler of future `front line' situations including emergency response and disaster relief. In such highly dynamic settings, coordination among team members is a critical success factor. The increasing sophistication of modern artificial intelligence (AI) has led to significantly improved perception, cognition, communication and action (PCCA) capabilities embodied in robots. However, many of the key technologies are `black box' in nature, making it hard to engineer robots that operate in a sufficiently transparent manner to their human collaborators. This work is a step towards analysing human expectations in terms of explainability in relation to task coordination. The setting is a simulation environment, TeamCollab~\cite{fusion}, in which humans and AI-equipped robots collaborate to clear an area of dangerous objects. The environment is designed to highlight PCCA capabilities; humans and robots are intended to work as peer agents, and inter-agent communication is a key factor in task success.

We analyse results from TeamCollab experiments with human participants to better understand what humans expect from their teammates in relation to explanation. We adopt a recent explainable AI (XAI) framework \cite{bertrand2023selective} that takes a dialogue-centric view of explanation, labelling the exchanged messages in terms of their relationship to elements of the framework. We show that there is a positive relationship between message exchange and team performance: volume of communication correlates with task success. 
This analysis seeks to answer the research question: What types of explanations do AI-equipped robots need based on human teammate expectations?
This is important, as AI agents that communicate effectively have been shown to earn more trust and cooperate better with humans \cite{Zhang2023}.

\section{RELATED WORK}

As AI systems will be required to perform in highly cognitively demanding environments, it has become critical to understand the communication mechanisms that would better support human-AI teams. Transparency and explainability are key components of situational awareness to allow agents to understand better the dynamically changing world they constantly perceive \cite{endsley2023supporting}. Thus, XAI aims to improve the understanding and interpretation of the decision processes and results of machine learning algorithms \cite{liao2020questioning, mueller2021principles} to support, e.g., human-robot teaming. Current approaches to XAI have mostly focused on static explainability, with a single message to cover, without input or user preferences involved in the process \cite{sokol2020one}. Due to the social nature of XAI, interactive explanations are gaining attention \cite{miller2019explanation}, as this involves an iterative process that considers user's information needs and is more similar to people's patterns on how explanations are expected to be provided \cite{bertrand2023selective, mueller2021principles}.
Some taxonomies have been proposed to explore the interactive aspect of XAI. Liao et. al \cite{liao2020questioning} present an XAI question bank framework with a set of prototypical questions that users may ask when requesting an explanation from an AI system. 
Authors in \cite{bertrand2023selective} synthesize 48 empirical studies to create a two-level taxonomy of interactive techniques in XAI based on their cognitive processes and tasks. 

During communication in critical timing scenarios, asking for an explanation may not always be explicit. The theory behind team communication in human groups identifies that a significant amount of communication goes through non-explicit channels, meaning that messages are interpreted in context, and there are additional communicative channels used as eye-gaze, gestures or non-verbal statements critical for performance that complement the message \cite{liang2019implicit}. Therefore, to better understand implicit explanation dialogues in context, we analyse the communication interaction of human-human data in a team environment and map awareness factors into a taxonomy of interactive techniques in XAI.



\section{METHODS}

\subsection{Experiment Design}

\begin{figure}
\centering
\includegraphics[width=0.45\textwidth]{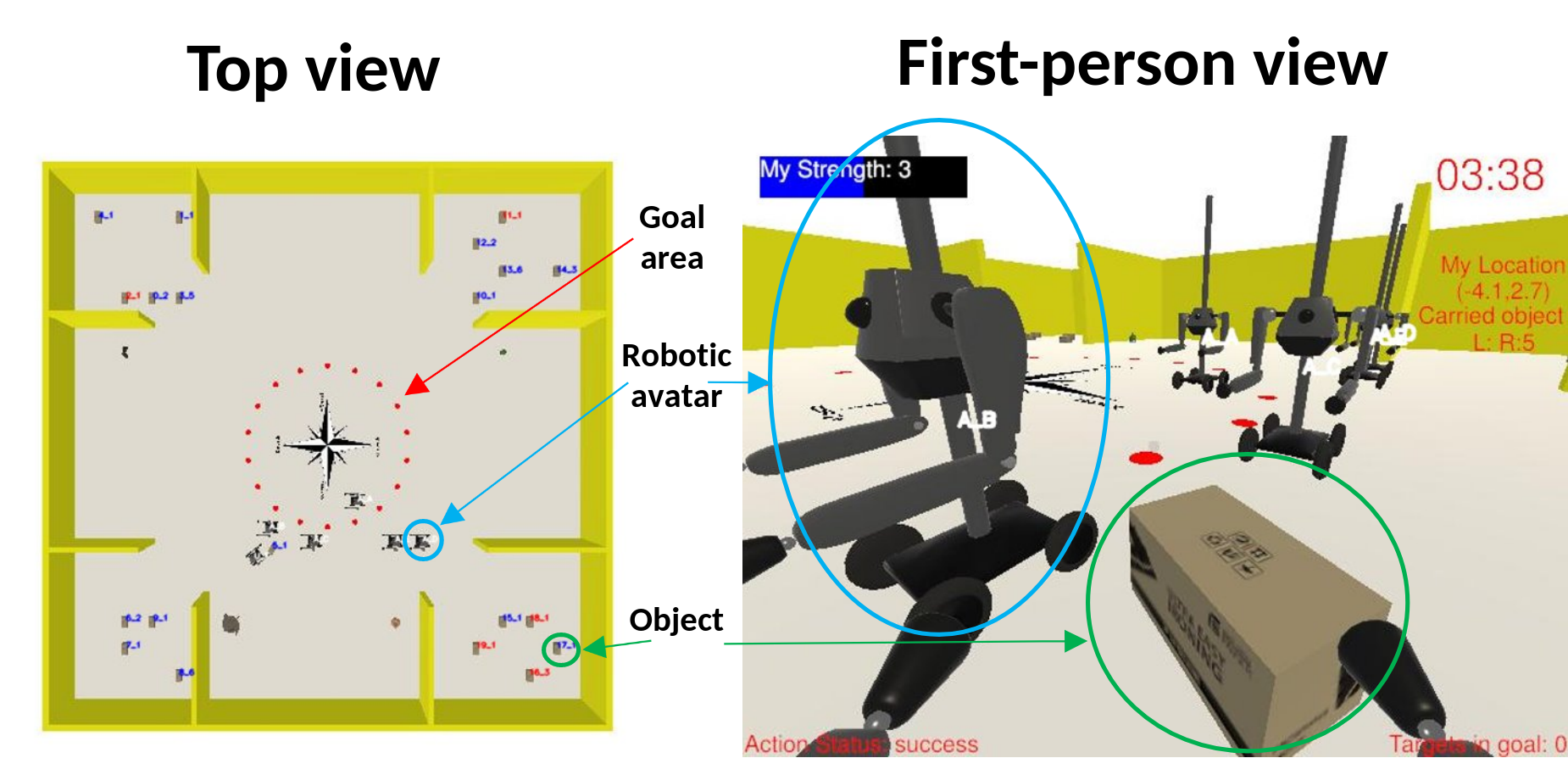}
\caption{\label{fig:platform} A top view and first-person view of the simulation environment TeamCollab~\cite{fusion}. Agents only have access to their avatar's first-person view.}
\end{figure}

The experiment was designed to run in TeamCollab~\cite{fusion}, a simulated environment based on the ThreeDWorld physics simulator \cite{Gan2021}. 
Participating agents are tasked with identifying dangerous objects distributed around a scenario and carrying them to a goal area, as shown in Fig. \ref{fig:platform}. Each object has two attributes: (i) weight, indicating the number of agents required to carry that object and (ii) danger, a boolean value indicating whether the object is dangerous or benign.
To obtain these attributes, each agent is equipped with a sensor. This sensor always accurately predicts the weight, but it can be inaccurate regarding the danger. The accuracy of the sensor regarding the danger varies between agents. Agents know how accurate their sensors are, but they can only learn if their prediction was correct for a given object by carrying it to the goal area. This is designed to encourage communication between agents to confirm which objects are actually dangerous. In addition, many of the objects are too heavy for a single agent to carry, requiring coordination to move them to the goal. The simulation thus emphasises perception, cognition, communication and action elements. It also facilitates collaboration between human and AI agents to achieve the shared goal of collecting all dangerous objects as fast as possible or, failing that, to collect as many dangerous objects as possible by a deadline. 

In our experiments, we used a deadline of 20 minutes, after which agents could no longer act. We considered groups of 2 to 4 human agents, with the intent of better understanding how we should implement robot agents so that they interact correctly with humans.

\subsection{Data Collection}

All participants were asked to connect to the simulation through their web browser and to avoid the use of any outside communication with other participants for the duration of the session. Participants were provided with a proximity-based text chat, where communications were broadcast to all agents within a range of 5 metres. The system recorded all communications, capturing the following parameters:
\begin{itemize}
    \item File \& session ID: identify each individual session;
    \item Timestamp: time in seconds when the message was sent;
    \item Sender: the agent that sent the message.
    \item Receivers: the agents that received the message;
    \item Message: the text that the agent sent.
\end{itemize}

We ran 20 sessions, resulting in 2,607 messages.

\subsection{Data Analysis}
Following a qualitative approach, we conducted a linguistic analysis of communication logs to identify explanation-related messages. Previous approaches have been followed to understand in-depth interactions that emerge from collaborative dialogues and spoken instructions \cite{fuentes2023roboclean,kullasaar2002developing}. We analyse the messages from the perspective of the recent XAI taxonomy presented by \cite{bertrand2023selective}. This framework was chosen because it offers a synthesis of findings from 48 empirical studies evaluating interactive explanations with human users.
The resulting taxonomy classifies XAI techniques according to the type of support interaction: (i) select, which allows users to choose the information they want to see, (ii) mutate, which considers hypotheses or different situations, and (iii) dialogue with, which provides interactivity. Each cognitive support type is divided into three task-oriented categories.
For our experiment, we focus on four categories: (i) select/clarify, which gives additional information on demand, (ii) mutate/simulate, which considers predictions for a given set of inputs, (iii) dialogue/progress, which guides the user through an explanation sequence and (iv) dialogue/answer, which gives feedback or edits explanation components. In our analysis, we are not only looking for communications that explicitly request explanation (e.g., questions asking, ``Why\ldots?'') but also identifying parts of the human-human dialogues where our participants are, in the judgement of the annotators, implicitly calling for explanation within the four sub-categories. The purpose here is to focus on the collaborative activities where robots will most need to be equipped to offer explanations in response to implicit as well as explicit requests.
Following an iterative process guided by a CodeBook \cite{decuir2011developing}, four researchers labelled 1,000 of the collected messages, which came from 15 different sessions.

\subsubsection{Procedure} Researchers used a 3-stage qualitative analysis to label the communication logs. First, 100 messages were individually labelled and checked for inter-code agreement, adjusting the CodeBook where necessary. Then, 150 more messages were labelled and discussed to ensure inter-code agreement. Finally, a total of 1,000 messages were labelled, including the re-labelled first 100 messages.

\subsubsection{CodeBook}
The team worked through the codes to label the messages, classifying and refining them until all the researchers agreed. The labels were then used as themes to classify the messages. Table \ref{tab:codebook-table} shows the final labels, along with a definition and the corresponding category from the taxonomy in \cite{bertrand2023selective} interpreted as noted above.

\begin{table*}[!t]
\caption{CodeBook description}
\label{tab:codebook-table}
    \begin{tabular}{|p{2cm}|p{2cm}|p{12.4cm}|}
    
    \hline
    Label & Category & Definition \\ \hline
    doing & select/clarify & The   message is either a question that can be interpreted as ``What are you doing?''   or a statement that can be interpreted as an answer to ``What are you doing?'' \\ \hline
    why & select/clarify & The message is either a question that asks for information to justify why some action has (or has not) been taken or a statement that provides such a justification. The action can be an observation or a decision. \\ \hline
    features & select/clarify & The message (question or statement) specifically refers to features of an object or agent. \\ \hline
    auto & select/clarify & Autocompleted messages used to share all sensed details for a given object via a button on the UI. \\ \hline
    dangerous & mutate/simulate & The message refers to actions / observations / decisions under an assumption that the object is dangerous. \\ \hline
    safe & mutate/simulate & As above but the assumption is the object is benign. \\ \hline
    make-safe & mutate/simulate & The message specifically refers to actions to make an object safe. \\ \hline
    order & dialogue/progress & The message is an order / instruction / command / tell. \\ \hline
    confirm & dialogue/answer & The   message is a confirmation. \\ \hline
    \end{tabular}
\end{table*}

\section{RESULTS}
\subsection{Communication Effect on Team Performance}

Before analyzing the messages sent, we first wanted to evaluate whether team communication had any impact on the performance of the team. For that purpose, in this section we compare the number of messages sent by the team as a whole with different performance indicators.

\begin{figure*} 
    \centering
  \subfloat[Number of objects collected\label{fig:performance_results:a}]{%
       \includegraphics[width=0.3\linewidth]{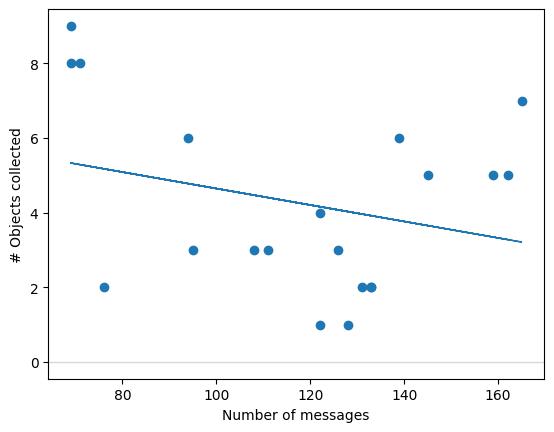}}
    \hfill
  \subfloat[Number of dangerous objects collected\label{fig:performance_results:b}]{%
        \includegraphics[width=0.3\linewidth]{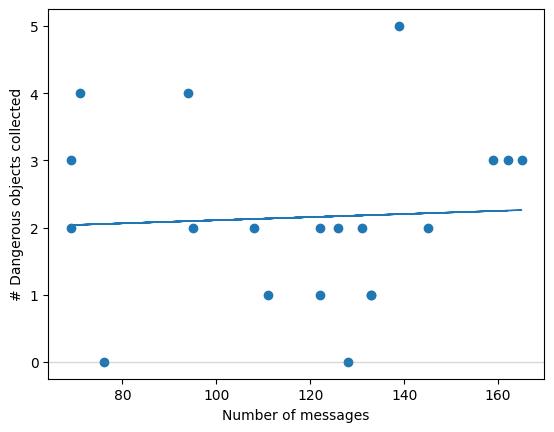}}
    \hfill
  \subfloat[Percentage of collected objects that are actually dangerous\label{fig:performance_results:c}]{%
        \includegraphics[width=0.3\linewidth]{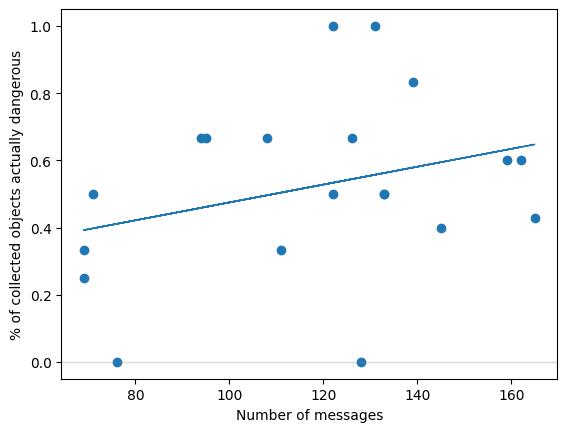}}
  \caption{Different evaluations of performance by the number of team messages sent.}
  \label{fig:performance_results} 
\end{figure*}

Fig.~\ref{fig:performance_results:a} shows that the total number of collected objects tends to have an inverse correlation to the number of messages sent by the team as a whole, with a Pearson Correlation Coefficient (PCC) of -0.28. Intuitively, this makes sense, as participants need to spend time typing the messages, which prevents them from taking other actions in the environment. As such, it might seem that sending messages is detrimental to the performance of the team as a whole. However, Fig.~\ref{fig:performance_results:b} shows that the number of actually dangerous objects collected by the teams does not seem to depend on the number of messages sent (PCC of 0.05). We believe this is thanks to the knowledge that each agent gains from communicating with the others, which allows everyone to compare the predictions from different sensors. This means agents can more accurately predict which objects are actually dangerous, leading to less wasted time on trips carrying benign objects. This is reflected in Fig.~\ref{fig:performance_results:c}, which shows that a higher percentage of the objects collected by more communicative teams are actually dangerous (PCC of 0.30).

As such, it seems that, while there may be some downsides to too much communication, the advantages of getting input from other teammates can make up for the time spent typing messages. 
It is also worth considering that in a real scenario, the disposal of dangerous objects might have a monetary or time cost beyond the time spent moving them to the goal area represented in the simulation. In such cases, being more accurate in identifying which objects are considered dangerous might have even further benefits.

\subsection{Understanding XAI-Related Language}

\begin{figure} 
    \centering
    \includegraphics[width=\linewidth]{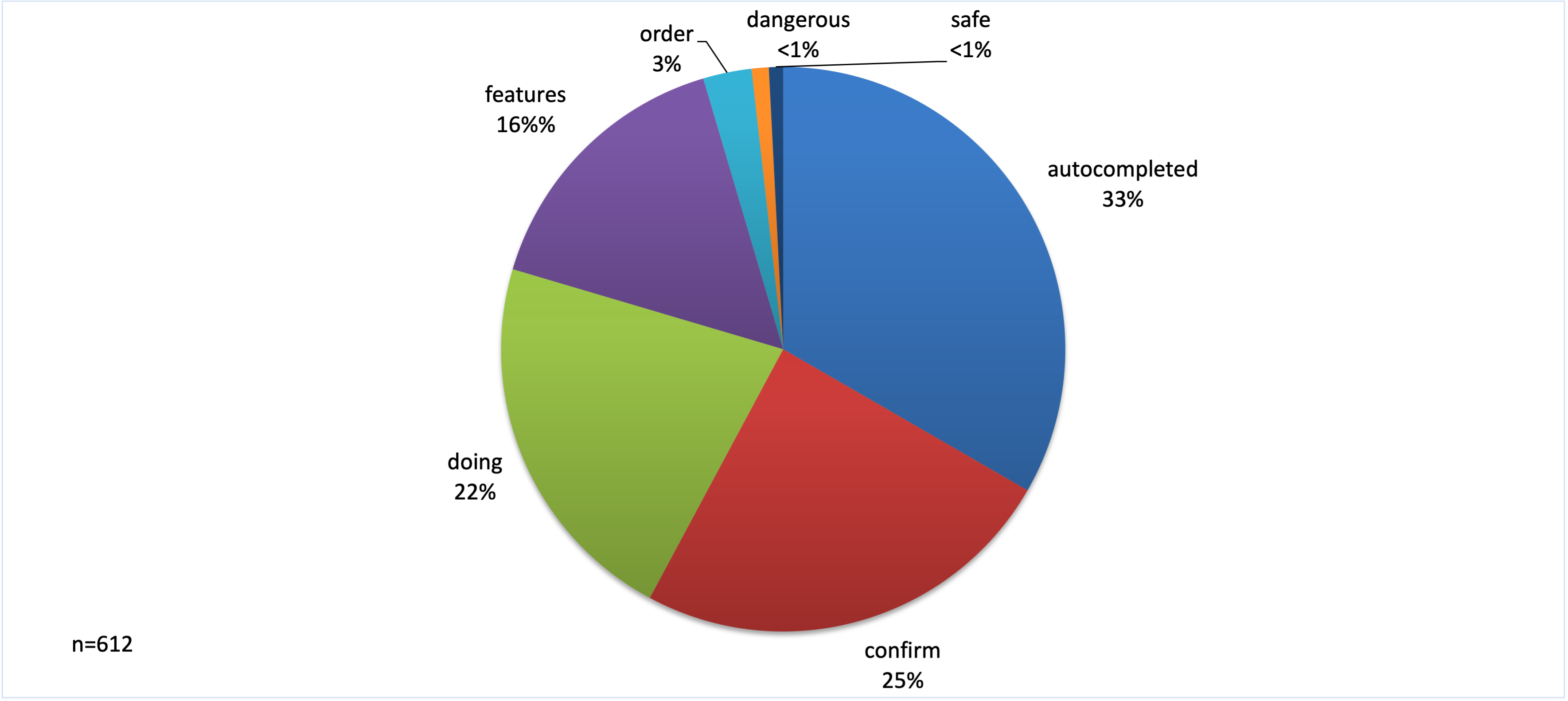}
    \caption{Majority Label Distribution ($\geq 3$ agreement level)}
    \label{fig:majority_label}
\end{figure}

We present the results of the qualitative linguist analysis conducted to understand what XAI language emerges from the communications. 
Four annotators categorised and labelled 1,000 messages with a good level of inter-annotator agreement (61\% of samples were labelled with at least three annotators agreeing on the label, and 5\% of samples were unclassified or with no similarities). Fig. \ref{fig:majority_label} shows the majority label distribution for the annotated messages. That is, the labels where at least 3 annotators agreed. The most frequent label is \textbf{auto}completed messages, which participants can easily send by pressing a UI button to share all sensed attributes for an object. An example of this type of message is \textit{``Object 18 (weight: 1) Last seen in (-7.5,-5.5) at 04:06 -Status Danger: dangerous,Prob. Correct: 72.5\%''.} The ease with which these messages can be generated likely had an influence on their frequency. The second most common label is \textbf{confirm}, for messages confirming communications, such as ``K sounds good'', and ``Ok 9''. This is followed by \textbf{doing}, for messages that either ask what actions are being taken or can be interpreted as answering such questions. Some examples are \textit{``I’ll explore in the next room'', ``do you want to pick up another one, or should we call it a day''.}
Messages discussing the \textbf{features} of an object are also common, including cases such as: \textit{``Nothing here for me. Only 2 but that one is too heavy''; 
``I see, that's high''.} All other labels are significantly less common, with \textbf{why} and \textbf{make-safe} having no cases of majority agreement. There also were no cases where a majority of annotators agreed on assigning multiple labels to a single message.

It is worth noting that most of the labelled messages fall into the select/clarify category (auto, doing, features), followed by dialogue/answer (confirm). 
Annotators also remarked that, while more traditional XAI questions such as \textit{``Why do you think object X is dangerous?''} do not tend to appear explicitly, participants did discuss why they thought an object was dangerous or not. This is likely why messages labelled with \textbf{auto} and \textbf{features} are quite frequent, as participants shared their opinions on how to act based on shared sensor outputs. Some messages from \textbf{confirm} were also part of such discussion, as conclusions were reached. Most of the other messages labelled \textbf{confirm} were in response to \textbf{doing} messages, which tended to be used to inform teammates of actions and coordinate.

\addtolength{\textheight}{-0.3cm}   

\section{DISCUSSION AND CONCLUSION}

In this work, we sought to gain a better understanding of what explanation capabilities are required of PCCA robots in team collaboration tasks in dynamic emergency response type settings. Our expectation that effective communication matters in such settings is borne out by data showing that the volume of dialogue correlates with the performance of the team. More specifically, we have shown that, while teams that communicate more tend to collect fewer objects, they are more accurate, with a higher percentage of collected objects being actually dangerous. As a result, we see that the implementation of TeamCollab is successful in rewarding teams that use effective communication while having the potential to penalise teams that communicate ineffectively.

Our analysis of the content of the communications revealed that a high proportion of messages discussed the attributes of the objects and what to do with them as a result. There was also a significant proportion of messages intended to coordinate the team, either as a high-level strategy or for a specific task. In terms of the PCCA capabilities, our analysis focused directly on \textit{communication}. The content of a high proportion of messages concerned \textit{perception} (\textbf{auto}/\textbf{features}) and \textit{action} (\textbf{doing}). We identify \textit{cognition} in the discussions deliberating whether an object is actually dangerous based on multiple sensor readings and in the messages trying to achieve team coordination.

These findings will inform the implementation of explanation-equipped AI agents that can effectively communicate with their human teammates. In particular, it will help us focus on the XAI capabilities that humans expect from their teammates. In immediate future work, we will extend the TeamCollab experiments to human-robot interaction and analyse whether there are any significant differences in how humans communicate with teammates depending on whether they are human or robot.
We would also like to explore any differences for larger teams and other tasks. 
Furthermore, we would like to run similar experiments in the real world to allow us to capture non-verbal communication aspects.





\section*{ACKNOWLEDGMENT}

The research reported in this paper was sponsored in part
by the DEVCOM Army Research Laboratory via cooperative
agreement W911NF2220243.
Any opinions, findings, and conclusions or recommendations expressed in this material are those of the author(s) and do not necessarily reflect the views of the United States government.



\bibliographystyle{IEEEtran}
\bibliography{mybibfile}

\end{document}